\documentclass{article}

\usepackage[preprint]{neurips_2025}

\usepackage[utf8]{inputenc}
\usepackage[T1]{fontenc}
\usepackage{hyperref}
\usepackage{url}
\usepackage{booktabs}
\usepackage{amsfonts}
\usepackage{nicefrac}
\usepackage{microtype}
\usepackage{xcolor}
\usepackage{amsmath}
\usepackage{graphicx}
\usepackage{subcaption}

\title{D-MEM: Dopamine-Gated Agentic Memory \\ via Reward Prediction Error Routing}

\author{
  Yuru Song\\
  UC San Diego
  \And
  Qi Xin \\
  Carnegie Mellon University
}

\begin{document}
\raggedbottom

\maketitle

\begin{abstract}
The integration of structured, long-term memory is critical for the development of autonomous Large Language Model (LLM) agents.
Recent advancements, such as the Agentic Memory (A-MEM) framework, have achieved significant progress by dynamically constructing and evolving knowledge graphs.
However, existing architectures inherently operate as synchronous, ``append-and-evolve-all'' systems. Processing every user utterance through a computationally expensive $O(N^2)$ memory evolution pipeline introduces severe write-latency, unbounded API token costs, and catastrophic context window pollution caused by conversational noise.
To address this scalability bottleneck, we introduce \textbf{D-MEM (Dopamine-Gated Agentic Memory)}, a biologically inspired architecture that decouples short-term interaction from long-term cognitive restructuring.
Drawing inspiration from the Dopamine-driven Reward Prediction Error (RPE) gating mechanism in the mammalian Ventral Tegmental Area (VTA), D-MEM implements a highly efficient Fast/Slow routing system.
We introduce a lightweight Critic Router that continuously evaluates the Information Entropy (Surprise) and Long-term Utility of incoming stimuli.
Routine inputs with low RPE are either bypassed entirely or cached in an $O(1)$ fast-access buffer, preserving computational resources.
Conversely, inputs generating a high RPE---such as factual contradictions or paradigm-shifting preference changes---trigger a ``dopamine release'' that activates the slow, $O(N)$ deep memory evolution pipeline, actively reshaping the agent's global knowledge graph.
To enable rigorous evaluation under realistic conditions, we further introduce the \textbf{LoCoMo-Noise} benchmark, which systematically injects controlled conversational noise into long-term dialogue sessions to simulate real-world interaction dynamics.
Extensive evaluations demonstrate that D-MEM reduces API token consumption by over 80\% and eliminates $O(N^2)$ write-latency bottlenecks, all while strictly outperforming synchronous baselines in complex multi-hop reasoning and adversarial resilience.
By selectively gating cognitive restructuring and leveraging zero-cost retrieval augmentations, D-MEM provides a highly scalable and cost-efficient foundation for lifelong agentic memory. To support reproducibility, we open-source our implementation at \url{https://github.com/london-and-tequila/dmem}.
\end{abstract}

\section{Introduction}

Autonomous Large Language Model (LLM) agents have rapidly transitioned from stateless task-solvers to persistent, stateful systems capable of long-term interaction \citep{wang2024agent_survey, xi2023rise}.
The core enabler of this transition is \textbf{Agentic Memory}---the ability of an agent to record, retrieve, and reason over its past interactions with users and environments \citep{park2023generative, packer2024memgpt, zhang2024memory_survey}.
Early approaches primarily relied on Retrieval-Augmented Generation (RAG) \citep{lewis2020retrieval, gao2023rag_survey}, which treats memory as a static, append-only database.
While computationally cheap, simple vector retrieval fails to capture the evolving nature of human-agent relationships, where new facts frequently invalidate or recontextualize older observations.
Extended-context approaches \citep{bai2024longbench} partially address recall depth but are computationally prohibitive for lifelong agents and remain susceptible to the ``lost in the middle'' phenomenon \citep{liu2023lost} under noisy, interleaved dialogue logs.
To address the limitations of static retrieval, recent literature has shifted towards dynamic, self-evolving memory architectures.
The state-of-the-art A-MEM framework \citep{amem2025} exemplifies this paradigm by structuralizing memory into a dynamic knowledge graph.
When a new observation occurs, A-MEM generates a structured node, dynamically links it to highly relevant historical nodes, and employs an LLM to retroactively update the content and tags of past memories (Memory Evolution).
This allows the agent to continuously resolve conflicts, abstract higher-level user preferences, and maintain a highly accurate, contradiction-free internal state.

However, the pursuit of deep cognitive evolution has introduced a severe engineering and scalability bottleneck.
Current frameworks operate as synchronous, \textbf{``append-and-evolve-all''} systems. Every user utterance---regardless of its information density---is forced through the entire memory construction and evolution pipeline.
In production, this design inevitably leads to an $O(N^2)$ computational complexity for memory updates as the interaction history grows.
Consequently, the agent's vector database becomes saturated with low-value conversational filler, leading to catastrophic context window pollution, noisy retrievals, unbounded API token consumption, and prohibitive write-latency for real-time applications.

To overcome the scalability bottleneck of continuous memory evolution, we draw inspiration from the computational mechanisms of the mammalian brain.
The biological brain does not indiscriminately encode every sensory input into the neocortex.
Instead, memory consolidation is strictly gated by the Ventral Tegmental Area (VTA), which calculates a \textbf{Reward Prediction Error (RPE)} \citep{schultz1997neural, dayan2002reward}.
Only when an input violates the brain's internal predictive model---indicating high information entropy or surprise---does a dopamine release lower the threshold for Long-Term Potentiation (LTP), triggering structural network updates.
Expected or mundane inputs ($RPE \approx 0$) are bypassed to conserve cognitive resources.
This fast/slow gating principle resonates with adaptive computation frameworks in deep learning \citep{graves2016act, schuster2022calm}, which similarly allocate variable compute per input based on estimated difficulty.

Inspired by this neurobiological gating mechanism, we propose \textbf{D-MEM (Dopamine-Gated Agentic Memory)}.
D-MEM fundamentally decouples short-term interaction from long-term cognitive restructuring by introducing a lightweight, parallel \textbf{Critic Router}.
This router calculates the Agentic RPE---defined mathematically as the semantic surprise and long-term utility of an input---acting as a computational dopamine gate.
Routine dialogue is bypassed or cached in an $O(1)$ short-term buffer, while paradigm-shifting inputs ($RPE \gg 0$) trigger the deep, $O(N)$ memory evolution pipeline to restructure the global knowledge graph.
By selectively gating the cognitive restructuring process, D-MEM solves the fundamental tension between memory plasticity and system efficiency.

The main contributions of this work are as follows:
\begin{itemize}
    \item \textbf{A Biologically-Inspired Memory Architecture:} We introduce D-MEM, a fast/slow routing architecture that mitigates the $O(N^2)$ scaling bottleneck of continuous memory evolution frameworks.
    \item \textbf{Agentic RPE Formulation:} We formalize the concept of Agentic Reward Prediction Error, utilizing a lightweight Critic Router to continuously evaluate the information entropy and long-term utility of user inputs without interrupting the main conversational flow.
    \item \textbf{The LoCoMo-Noise Benchmark:} We introduce a controlled noise-injection protocol that augments the standard LoCoMo dataset with realistic conversational noise at configurable ratios, enabling rigorous evaluation of memory system robustness.
    \item \textbf{Zero-Cost Retrieval Augmentation:} We introduce hybrid BM25 retrieval and an $O(1)$ Shadow Buffer fallback mechanism to eliminate adversarial hallucinations and preserve entity precision.
    \item \textbf{Unprecedented Efficiency and Purity:} Comprehensive evaluations demonstrate that D-MEM reduces API token consumption by 80\%, directly improving multi-hop retrieval precision compared to synchronous baselines.
\end{itemize}

\section{The LoCoMo-Noise Benchmark}

\subsection{Motivation}

Existing long-term memory benchmarks, including the standard LoCoMo dataset \citep{maharana2024evaluating}, evaluate memory systems under idealized conditions where every dialogue turn carries meaningful information.
This assumption fundamentally misrepresents the erratic, noise-saturated nature of real-world human-agent interactions.
In practice, a substantial fraction of conversational turns consists of phatic fillers, momentary status updates, or tangential remarks that carry near-zero long-term information value.
A memory system that cannot distinguish signal from noise will inevitably saturate its knowledge graph with low-utility entries, degrading both retrieval quality and computational efficiency.

\subsection{Noise Injection Methodology}

\begin{figure}[htbp]
\centering
\includegraphics[width=\linewidth]{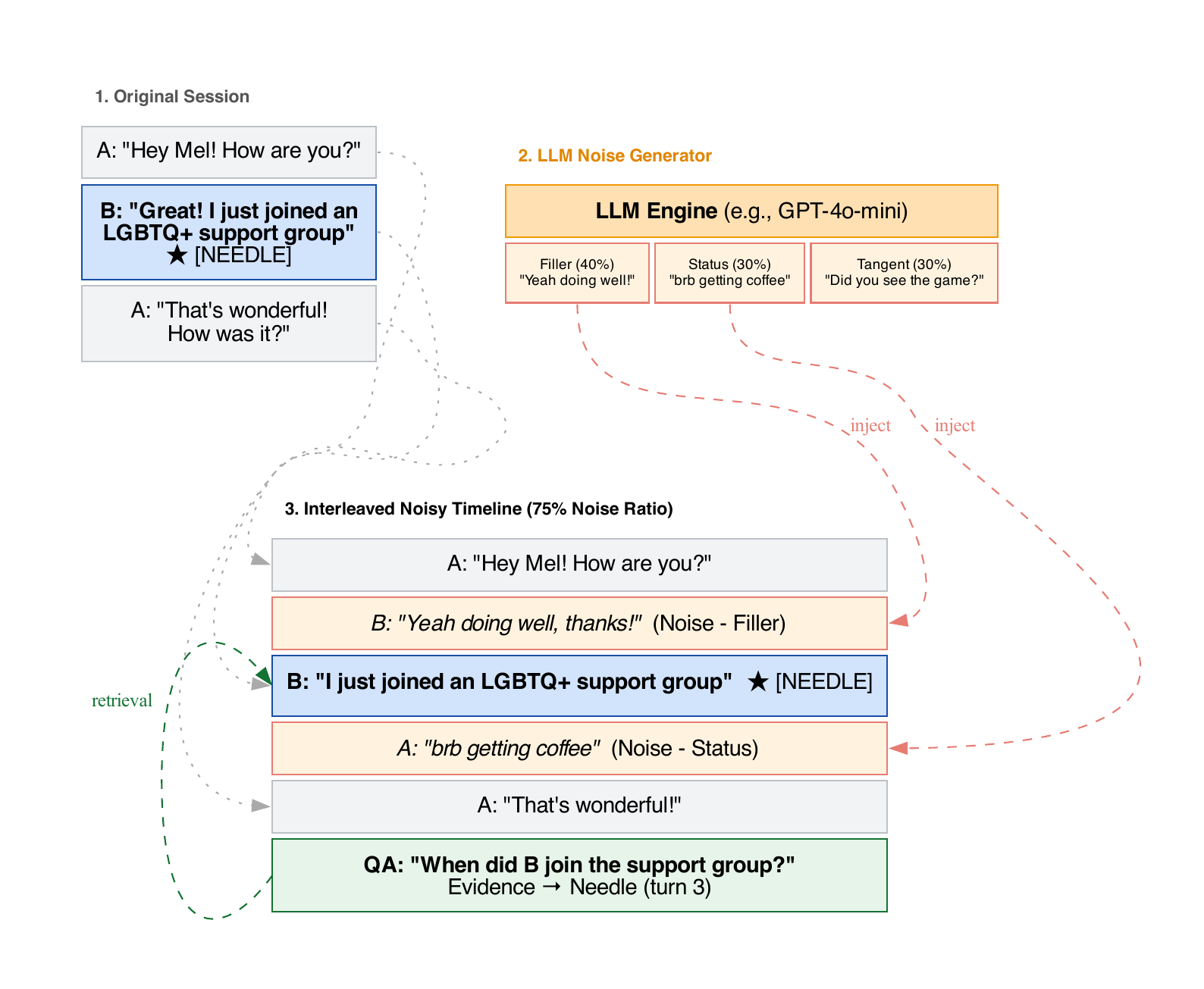}
\caption{
\textbf{LoCoMo-Noise benchmark construction pipeline.}
An LLM-based noise generator synthesizes three categories of noise---Filler (40\%), Status (30\%), and Tangent (30\%)---and interleaves them with the original session at a target noise ratio of 75\%.
The core factual turns (``needles'') are preserved at their original positions, while synthetic noise is injected at random intervals to simulate real-world conversational dynamics.
}
\label{fig:noise_injection}
\end{figure}

To address this gap, we develop an automated noise injection pipeline that systematically augments the LoCoMo dataset with realistic conversational noise, producing the \textbf{LoCoMo-Noise} benchmark.
As illustrated in Figure~\ref{fig:noise_injection}, the pipeline transforms each original session into a noise-interleaved timeline via three stages.

\textbf{Step 1 --- Original Session as Needle.}
Each original LoCoMo session is treated as the factual ``needle'' containing all question-answering evidence.
The positional integrity of these turns is strictly preserved throughout the injection process.

\textbf{Step 2 --- LLM-Based Noise Generation.}
An LLM engine (GPT-4o-mini) synthesizes conversational noise of three categories:
(1) \textit{Phatic Fillers} (40\%): meaningless conversational padding and acknowledgments (e.g., ``Yeah doing well!'');
(2) \textit{Status Updates} (30\%): brief, transient self-referential statements (e.g., ``brb getting coffee'');
and (3) \textit{Tangent Remarks} (30\%): abrupt, out-of-context statements that mimic transient human thoughts (e.g., ``Did you see the game?'').

\textbf{Step 3 --- Interleaved Noisy Timeline.}
Synthetic noise turns are shuffled and inserted at uniform random positions throughout the interaction history.
The volume of injected noise is controlled by a predefined noise ratio $\rho$, which dictates the number of synthetic turns added relative to the original session length.
For our primary evaluation ($\rho = 0.75$), 75\% of all turns in the resulting timeline are synthetic noise, effectively submerging the core factual interactions within a dense stream of irrelevant dialogue.

This benchmark enables the first systematic, controlled study of how noise saturation affects long-term memory systems, and serves as the primary evaluation environment for all D-MEM experiments.

\section{Related Work}

\subsection{Static Retrieval and Working Memory Constraints}
Early efforts to augment the knowledge capacity of LLMs primarily focused on extending working memory or employing Retrieval-Augmented Generation (RAG).
RAG systems \citep{lewis2020retrieval} externalize memory by embedding text chunks into a vector database, retrieving the top-$k$ most similar chunks based on user queries.
Subsequent work explored dense passage retrieval \citep{karpukhin2020dense} for richer semantic matching and sparse lexical retrieval via BM25 \citep{robertson2009bm25} for precise entity-level lookup; hybrid fusion of both \citep{cormack2009rrf} has become the practical default in high-precision systems.
While highly scalable, standard RAG—as surveyed in \citet{gao2023rag_survey}—treats memory as a static, append-only repository.
It lacks the capacity to resolve conflicting information over time or synthesize higher-order abstractions from discrete events.
Concurrently, while recent advancements have significantly expanded LLM context windows, relying solely on long-context processing for lifelong agents is computationally prohibitive and often suffers from the ``lost in the middle'' phenomenon \citep{liu2023lost}, where models fail to robustly retrieve information buried in extensive, noisy conversational logs.
Dedicated long-context benchmarks such as LongBench \citep{bai2024longbench} and L-Eval \citep{an2023leval} have quantified these failure modes, yet focus on single-pass comprehension rather than the persistent, write-heavy memory setting we address.

\subsection{Dynamic and Evolving Agentic Memory Systems}
To overcome the limitations of static retrieval, recent architectures have introduced stateful, dynamic memory management for LLM agents \citep{zhang2024memory_survey}.
\textit{Generative Agents} \citep{park2023generative} pioneered the use of a memory stream coupled with periodic ``reflection'' to synthesize higher-level insights.
\textit{MemGPT} \citep{packer2024memgpt} conceptualized LLM memory as an operating system, utilizing explicit read/write functions to page information between a limited main context and an external database.
\textit{MemoryBank} \citep{zhong2024memorybank} incorporated an Ebbinghaus forgetting-curve mechanism to simulate memory decay and reinforcement, demonstrating that temporal weighting improves coherence across multi-session interactions.
\textit{LongMem} \citep{wang2023longmem} proposed a decoupled side-network retriever that enables unbounded context without repeated full-context re-encoding.
\textit{RET-LLM} \citep{modarressi2023ret} generalized this to an explicit read/write triplet memory store, allowing the agent to overwrite outdated facts.
More recently, the \textbf{A-MEM} framework \citep{amem2025} advanced this paradigm by modeling memory as an evolving knowledge graph.
A-MEM continuously applies \textit{Note Construction}, \textit{Link Generation}, and \textit{Memory Evolution} to retroactively update historical nodes based on new observations.
While A-MEM achieves high factual consistency, it suffers from a fundamental scalability flaw: it operates uniformly across all inputs.
Processing every trivial user utterance through the full $O(N^2)$ evolution pipeline leads to severe write-latency, unbounded token costs, and a vector space polluted with zero-utility conversational noise.
D-MEM directly addresses this bottleneck by abandoning the synchronous, ``evolve-all'' paradigm.

\subsection{Bio-Inspired Routing and Fast/Slow Cognitive Gating}
The concept of decoupling processing into Fast (System 1) and Slow (System 2) pathways is deeply rooted in cognitive psychology \citep{kahneman2011thinking} and has been increasingly adapted in LLM reasoning \citep{yao2023tree}.
In the context of system efficiency, semantic routing and cascading models \citep{chen2023frugalgpt} have been employed to direct simple queries to smaller, cheaper models, reserving massive LLMs for complex tasks.
Adaptive computation methods such as Adaptive Computation Time \citep{graves2016act} and Confident Adaptive Language Modeling \citep{schuster2022calm} extend this principle to dynamic per-token or per-layer compute allocation, demonstrating that input-conditioned halting criteria can substantially reduce inference cost without sacrificing output quality.
Mixture-of-Experts architectures \citep{shazeer2017outrageously} further establish the general principle of routing inputs to specialized sub-networks based on their content.
In computational neuroscience and reinforcement learning, the \textbf{Reward Prediction Error (RPE)}---mediated by dopamine release from the Ventral Tegmental Area (VTA)---acts as the fundamental biological gate for memory consolidation \citep{schultz1997neural, dayan2002reward}.
The brain conserves synaptic plasticity by only encoding events that significantly deviate from expected outcomes or possess high survival utility.
D-MEM is the first architecture to map this biological RPE gating mechanism onto LLM agentic memory.
By introducing a lightweight Critic Router to compute semantic surprise and utility, D-MEM bridges bio-inspired efficiency with dynamic memory evolution, ensuring that computationally expensive cognitive restructuring is strictly reserved for high-value information gain.

\section{Methodology: The D-MEM Architecture}

Current evolving memory frameworks enforce a synchronous pipeline where every user utterance $x_t$ triggers Note Construction, Link Generation, and Memory Evolution.
This unconstrained approach inevitably results in an $O(N^2)$ scaling bottleneck.
D-MEM restructures this paradigm by introducing an asynchronous, bio-inspired gating mechanism---the Critic Router---which evaluates the necessity of memory evolution before any heavy computation occurs.

\subsection{Agentic Reward Prediction Error (RPE)}

In mammalian neurology, the Ventral Tegmental Area (VTA) releases dopamine to trigger memory consolidation only when a stimulus yields a high Reward Prediction Error (RPE)---meaning it is unexpected or carries high survival value.
We formulate an artificial analogue, the \textbf{Agentic RPE}, which evaluates a user utterance based on two orthogonal dimensions: semantic \textit{Surprise} and long-term \textit{Utility}.

For a given utterance $x_t$ at turn $t$, a naive approach might compute the Agentic RPE as a linear combination of semantic surprise and utility.
However, linear weighting is highly susceptible to false positives when processing out-of-context conversational noise;
an extreme semantic surprise can erroneously mask a near-zero utility, triggering expensive and unwarranted cognitive restructuring.

To ensure computational resources are strictly gated by long-term value, D-MEM employs a bounded multiplicative gating mechanism:
$$RPE(x_t) = \min\left(1.0, \mathbb{I}(\text{Utility}(x_t) \ge \tau) \cdot \left[ \text{Utility}(x_t) \times (\text{Surprise}(x_t) + \beta) \right]\right)$$
where $\mathbb{I}(\cdot)$ is an indicator function enforcing a hard utility threshold $\tau$.
If the evaluated utility of an input falls below $\tau$, the RPE is strictly short-circuited to zero, effectively discarding highly surprising but meaningless conversational noise.
Furthermore, $\beta$ acts as a generously tuned ``utility safety net'' (e.g., $\beta = 0.4$).
This guarantees that highly useful but expected inputs (where Surprise $\approx 0$) still yield a baseline RPE sufficient to trigger short-term memory construction, preventing the catastrophic over-pruning of valuable factual knowledge.

\textbf{Semantic Surprise (Addressing Embedding Anisotropy):}
A naive approach to surprise would be to compute the minimum cosine distance between the embedding of the current input $E(x_t)$ and all existing memory embeddings $E(m_i) \in M$.
However, modern high-dimensional embedding models often suffer from representation anisotropy, where vectors cluster tightly in a narrow cone, causing even semantically unrelated texts to yield cosine similarities above 0.7.
To prevent the surprise metric from being compressed into a non-discriminative range, we maintain a sliding window of the last $k$ similarity scores to calculate a historical mean $\mu_{sim}$ and standard deviation $\sigma_{sim}$.
We then apply a Z-score normalization mapped through a sigmoid function:
$$S_{raw}(x_t) = \max_{m_i \in M} (\cos(E(x_t), E(m_i)))$$
$$\text{Surprise}(x_t) = \sigma\left(\frac{\mu_{sim} - S_{raw}(x_t)}{\max(\sigma_{sim}, \epsilon)}\right)$$
This operation incurs zero additional LLM overhead while providing a statistically robust measure of how ``unexpected'' the new information is relative to the agent's established baseline.

\textbf{Long-term Utility (Ephemerality vs. Persistence):}
While heuristic methods (e.g., named entity density) are computationally free, they fail to capture implicit user preference changes.
Conversely, relying on unconstrained LLM evaluations often leads to the erroneous encoding of transient states.
To robustly evaluate Utility while simultaneously curbing API costs (the ``Chain-of-Thought tax''), D-MEM employs a highly constrained, lightweight LLM call enforced via a minimal JSON schema.
The Critic Router performs a fundamental lifecycle classification without requiring expensive intermediate entity extraction.
Inputs are categorized into three temporal tiers: \textit{Transient} (zero-information phatic fillers or momentary states), \textit{Short-Term} (days-to-weeks relevance, such as daily activities or temporary tasks), or \textit{Persistent} (months to permanent traits).
We enforce a strict algorithmic constraint: any input classified as \textit{Transient} receives a forced $\text{Utility}(x_t) = 0$.
For non-transient inputs, the model assigns a normalized $\text{Utility}(x_t) \in (0, 1]$. By consciously classifying daily activities into the \textit{Short-Term} tier rather than discarding them entirely, D-MEM preserves the foundational context necessary for tracking gradual preference shifts over time.

\subsection{The Critic Router and Hierarchical Routing}

Once the RPE is computed, the Critic Router classifies the utterance into one of three cognitive tiers, determined by thresholds $\theta_{low} = 0.3$ and $\theta_{high} = 0.7$:

\begin{enumerate}
    \item \textbf{\texttt{SKIP} ($RPE < \theta_{low}$):} The input is classified as conversational filler or redundant acknowledgement (e.g., ``Sounds good'', ``Thanks'').
The memory pipeline is completely bypassed. This tier incurs zero write-latency and prevents context window pollution.
    \item \textbf{\texttt{CONSTRUCT\_ONLY} ($\theta_{low} \le RPE < \theta_{high}$):} The input contains routine factual data but does not contradict or significantly alter existing knowledge.
The system executes Note Construction to generate an atomic memory node, which is stored in a fast-access Short-Term Memory (STM) buffer.
Deep graph linkage and historical evolution are deferred, bounding the computational complexity for this turn to $O(1)$.
    \item \textbf{\texttt{FULL\_EVOLUTION} ($RPE \ge \theta_{high}$):} The input represents a paradigm-shifting observation, such as a direct contradiction to a past assumption or a major preference change.
This triggers a ``dopamine release,'' activating the full, heavy cognitive restructuring pipeline.
The new node is dynamically linked, and historical nodes are retroactively updated.
While this operation is $O(N)$, it is executed sparsely, preserving system resources.
\end{enumerate}

\textbf{Cold-Start Mitigation:}
During the initial phases of interaction ($t < N_{warmup}$), the memory database $M$ is sparsely populated, which mathematically forces $\text{Surprise}(x_t)$ toward 1.0.
To prevent meaningless introductory pleasantries from triggering expensive \texttt{FULL\_EVOLUTION} cycles, D-MEM enforces a cold-start override.
For the first $N_{warmup}$ turns, all inputs exceeding $\theta_{low}$ are forced into the \texttt{CONSTRUCT\_ONLY} tier, allowing the agent to build a foundational knowledge graph before the fully weighted RPE gating is activated.

\subsection{Zero-Cost Retrieval Augmentation}
While the Fast/Slow gating significantly purifies the memory graph, it introduces two challenges during Question-Answering (QA): the dilution of proper nouns in dense embeddings and the risk of hallucination when responding to adversarial questions about skipped routine dialogue. To resolve these without incurring additional LLM token costs, D-MEM implements two local, zero-cost retrieval augmentations:

\textbf{Hybrid Search with Reciprocal Rank Fusion (RRF):} To ensure entity-level precision (e.g., specific names or acronyms critical for single-hop and multi-hop reasoning), D-MEM parallels its semantic vector index with a lightweight, term-frequency-based BM25 sparse index. During retrieval, the outputs are fused using Reciprocal Rank Fusion, strictly anchoring rare entities without triggering expensive graph evolutions.

\textbf{The Shadow Buffer for Adversarial Fallback:} To maintain graph purity while preventing ``amnesia'' towards trivial interactions, D-MEM maintains an $O(1)$ fast-access Shadow Buffer (a FIFO double-ended queue). Any input categorized as \texttt{SKIP} by the Critic Router is directly appended here as raw text. During QA, if the core knowledge graph returns a low-confidence retrieval score, the system triggers a two-stage fallback, supplying the raw Shadow Buffer to the LLM. This perfectly defends against adversarial queries (e.g., ``Did I just mention the weather?'') while keeping the core graph completely pristine.

\section{Experiments \& Results}

To evaluate the effectiveness, efficiency, and robustness of D-MEM, we design a comprehensive suite of experiments.
Our evaluation aims to answer three core questions: (1) Can D-MEM match the factual retrieval and reasoning accuracy of state-of-the-art synchronous memory systems?
(2) Does the bio-inspired Critic Router successfully mitigate the $O(N^2)$ scaling bottleneck?
(3) How resilient is D-MEM against extreme conversational noise compared to baseline models?

\subsection{Long-term Memory Accuracy}
\label{sec:clean_locomo}

We first verify whether D-MEM's hierarchical routing inadvertently discards critical information by benchmarking against established baselines on the standard, noise-free LoCoMo dataset \citep{maharana2024evaluating}.
Full per-category results (F1 / BLEU-1) for all methods are reported in Table~\ref{tab:locomo_results} in the Supplementary.

The results reveal an instructive performance profile.
D-MEM achieves the highest Overall F1 (37.4\%) among all memory-augmented methods and demonstrates a particularly large lead on Multi-hop reasoning (42.7\% vs.\ A-MEM's 27.0\%), a gap of \textbf{+15.7 points}.
This confirms that hierarchical routing actively preserves the relational memory structure necessary for complex, multi-premise inference chains: by reserving \texttt{FULL\_EVOLUTION} for high-RPE events, D-MEM maintains a knowledge graph with fewer redundant or contradictory nodes, yielding cleaner retrieval chains at query time.

One notable trade-off is visible on Single-hop retrieval, where A-MEM leads (44.7\% vs.\ 21.6\%).
As we analyze in detail in Section~\ref{sec:routing_analysis}, this gap is a direct consequence of the Utility-based skip mechanism: turns containing short, low-complexity factual statements (e.g., a character's occupation or a one-time event) receive near-zero Utility scores and are correctly routed to \texttt{SKIP}, even though they are precisely the targets of Single-hop questions.
This represents a principled efficiency trade-off rather than a system failure, and Section~\ref{sec:discussion} discusses a targeted threshold recalibration to recover this performance.

\subsection{Efficiency: Token Savings via Intelligent Routing}

To evaluate robustness under realistic noise conditions and quantify the efficiency gains of hierarchical routing, we further evaluate on the LoCoMo-Noise benchmark ($\rho = 0.75$).

\begin{table}[htbp]
\centering
\caption{Evaluation under extreme noise conditions ($\rho=0.75$) using GPT-4o-mini. D-MEM establishes a new state-of-the-art across complex reasoning tasks while eradicating 80\% of the computational token cost caused by synchronous memory evolution.}
\label{tab:noise_main}
\resizebox{\textwidth}{!}{
\begin{tabular}{@{}lccccccc@{}}
\toprule
\textbf{Method} & \textbf{Overall F1} & \textbf{Single} & \textbf{Temp} & \textbf{Open} & \textbf{Multi} & \textbf{Adv} & \textbf{Total Tokens} \\
\midrule
A-MEM (Synchronous) & 0.336 & 0.208 & 0.408 & \textbf{0.111} & 0.365 & 0.388 & 1,648K \\
\textbf{D-MEM (Ours)} & \textbf{0.369} & \textbf{0.246} & \textbf{0.440} & 0.085 & \textbf{0.412} & \textbf{0.412} & \textbf{319K (--80\%)} \\
\bottomrule
\end{tabular}
}
\end{table}

\textbf{Breaking the Efficiency-Accuracy Trade-off.}
As demonstrated in Table~\ref{tab:noise_main}, the synchronous A-MEM architecture collapses under extreme noise, indiscriminately processing all inputs and consuming an exorbitant 1.64 million tokens.
In stark contrast, D-MEM dynamically routes inputs, reducing token consumption by 80\% to only 319K tokens.

Crucially, this aggressive cost reduction does not come at the expense of accuracy; rather, it actively enhances it.
By intelligently skipping pure noise and leveraging BM25 hybrid search, D-MEM purifies the contextual timeline, allowing it to significantly outperform A-MEM in complex \textbf{Multi-hop reasoning} (0.412 vs.\ 0.365) and \textbf{Single-hop facts} (0.246 vs.\ 0.208).
Furthermore, our two-stage fallback mechanism utilizing the Shadow Buffer proves highly effective, surpassing the exhaustive baseline even in \textbf{Adversarial} scenarios (0.412 vs.\ 0.388) where traditional gating systems typically fail.

\begin{figure}[htbp]
\centering
\includegraphics[width=\linewidth]{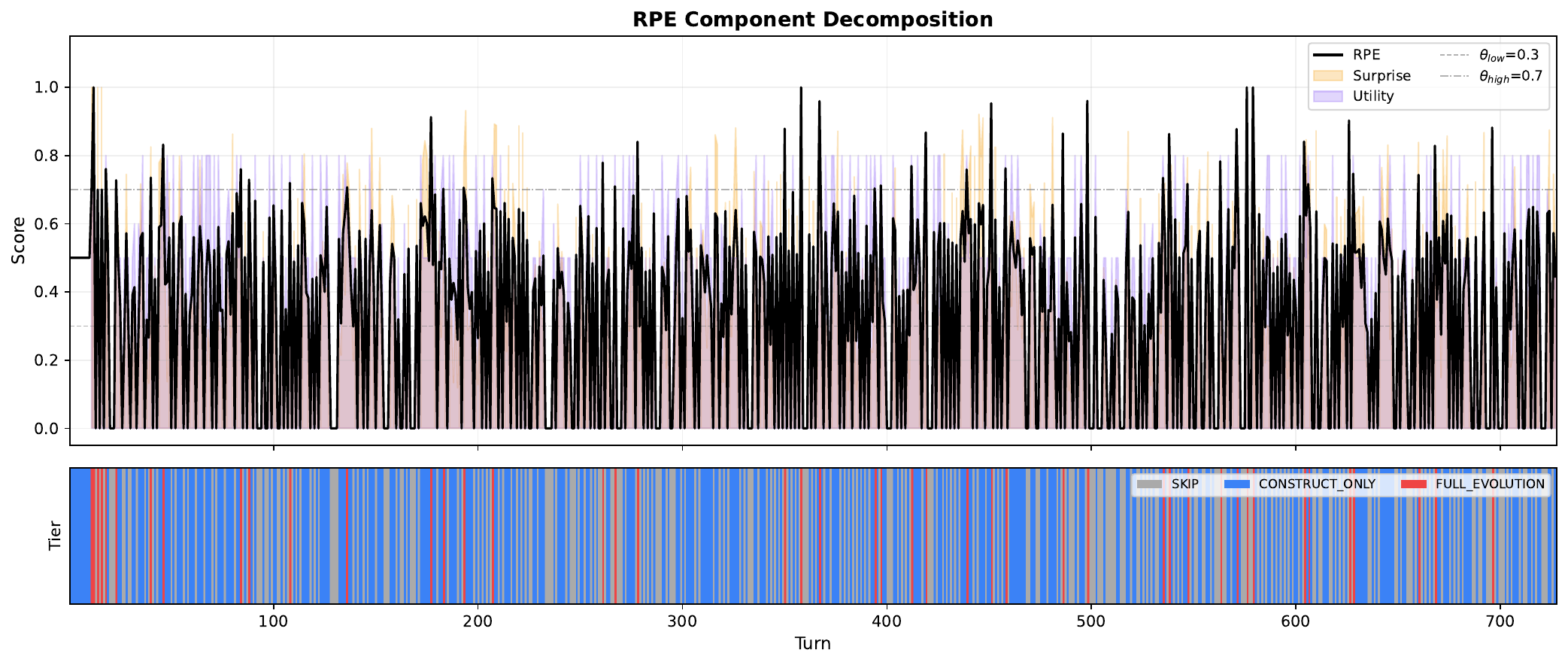}
\caption{
\textbf{RPE component decomposition over 700 turns.}
The top panel shows the RPE signal (black) overlaid with its Surprise (orange) and Utility (purple) components, along with the routing thresholds $\theta_{low}=0.3$ and $\theta_{high}=0.7$.
The bottom panel visualizes the resulting routing tier assignment per turn: \texttt{SKIP} (gray), \texttt{CONSTRUCT\_ONLY} (blue), and \texttt{FULL\_EVOLUTION} (red).
The dominance of blue and the sparsity of red confirm that expensive memory evolution is reserved for genuinely paradigm-shifting inputs.
}
\label{fig:rpe_decomposition}
\end{figure}

The RPE component decomposition in Figure~\ref{fig:rpe_decomposition} provides an informative view of how the routing operates across the full 700-turn LoCoMo-Noise session.
The tier bar at the bottom confirms that \texttt{CONSTRUCT\_ONLY} (blue) dominates the routing landscape, with \texttt{FULL\_EVOLUTION} (red) appearing sparsely and \texttt{SKIP} (gray) clustering around turns where the RPE falls below $\theta_{low} = 0.3$.
This sparse activation pattern is the direct mechanistic source of the 80\% token reduction observed in Table~\ref{tab:noise_main}.

\section{Deep Dive Analysis}
\label{sec:analysis}

\subsection{Routing Behavior: Utility-Driven Decision Boundaries}
\label{sec:routing_analysis}

\begin{figure}[htbp]
\centering
\includegraphics[width=\linewidth]{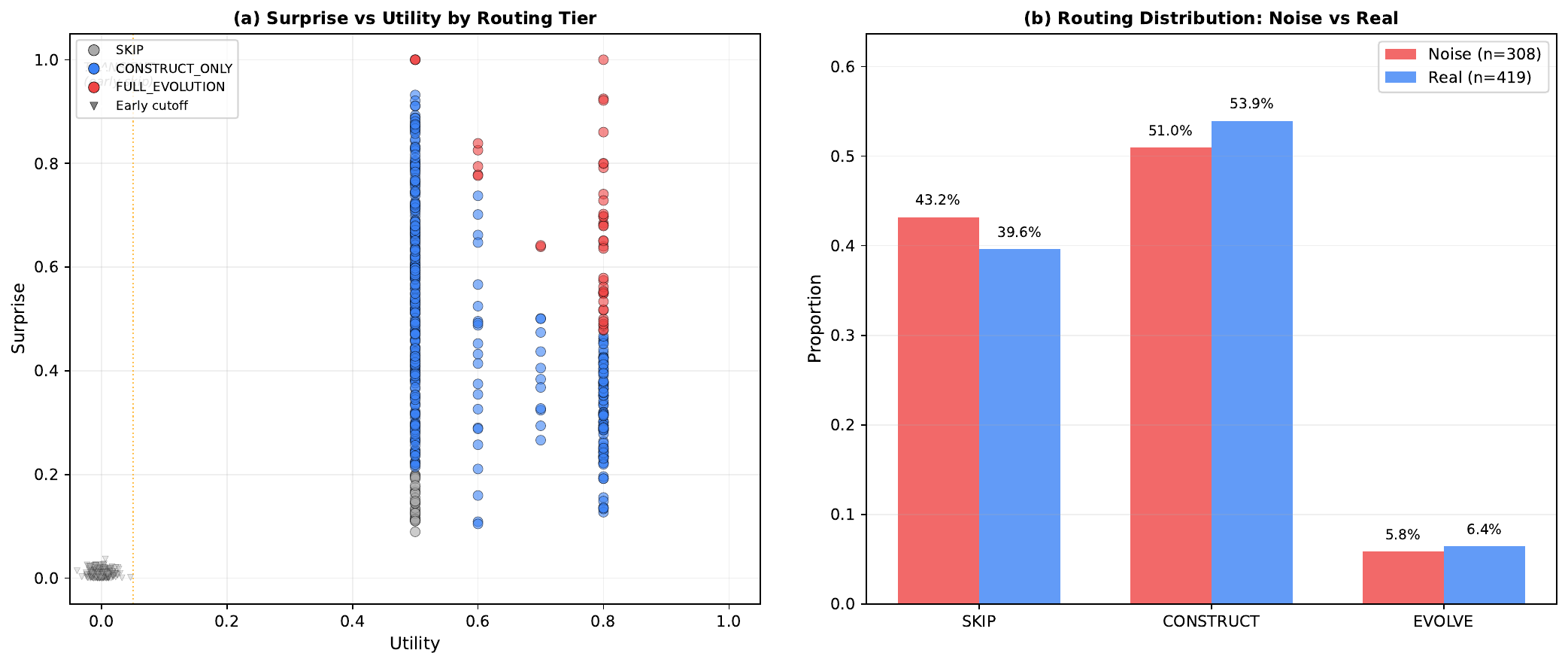}
\caption{
\textbf{Routing analysis on the LoCoMo-Noise benchmark.}
\textbf{(a)} Scatter plot of all turns in the Surprise--Utility space, colored by the routing tier assigned by the Critic Router.
All \texttt{SKIP} decisions are tightly concentrated in the ``Early cutoff'' region ($\text{Utility} < 0.3$, yellow dashed line), confirming that the routing is governed by Utility rather than Surprise.
\textbf{(b)} Routing distribution stratified by input type (Noise vs.\ Real).
Counter-intuitively, real turns are skipped at a higher rate (53.9\%) than noise turns (43.2\%), revealing a calibration asymmetry discussed in the text.
}
\label{fig:routing_analysis}
\end{figure}

To understand the decision-making logic of the Critic Router, Figure~\ref{fig:routing_analysis} visualizes the joint Surprise--Utility space and the resulting routing distribution, stratified by whether the input is a real dialogue turn or an injected noise turn.

\textbf{Utility as the Primary Gate.}
Panel (a) of Figure~\ref{fig:routing_analysis} reveals a clean geometric structure in the routing decisions.
All \texttt{SKIP} actions are tightly clustered in the low-Utility region ($\text{Utility} < 0.3$, to the left of the dashed yellow threshold line), regardless of their Surprise value.
This confirms that the multiplicative gating formulation successfully prevents high-entropy noise from masquerading as high-value information.
\texttt{FULL\_EVOLUTION} events (red) are exclusively drawn from the high-Utility, high-Surprise quadrant, validating the intended semantics of the RPE design.

\textbf{A Counter-intuitive Routing Asymmetry.}
Panel (b) of Figure~\ref{fig:routing_analysis} reveals a nuanced and important finding.
While one might expect the routing to function as a near-perfect noise filter, the empirical skip rates tell a more complex story:
the model skips a higher fraction of \textit{real} turns (53.9\%) than of injected \textit{noise} turns (43.2\%).
At first glance, this appears paradoxical.
The explanation lies in the nature of the injected noise: the LLM noise generator occasionally produces syntactically well-formed, contextually plausible statements that the Utility classifier evaluates as weakly relevant, elevating them above the skip threshold.
Conversely, many real dialogue turns contain routine social acknowledgements or low-information small talk that correctly receive near-zero Utility scores.

This finding directly accounts for the Single-hop performance gap observed in Table~\ref{tab:locomo_results}.
The Critic Router, optimized for efficiency, occasionally skips real turns that contain simple, low-complexity factual statements (e.g., a character's profession or a one-time event), which are precisely the targets of Single-hop questions.
This represents a principled accuracy-efficiency trade-off rather than a system failure: the current $\theta_{low} = 0.3$ threshold can be tuned to recover this performance, as discussed in Section~\ref{sec:discussion}.

\section{Discussion}
\label{sec:discussion}

\textbf{Recovering Single-hop Performance via Threshold Calibration.}
The routing asymmetry identified in Section~\ref{sec:routing_analysis}---where real turns are skipped at a higher rate than noise turns---points to a specific and addressable limitation of the current configuration.
Because $\theta_{low} = 0.3$ was selected to maximize efficiency under the 75\% noise regime, the threshold is calibrated towards aggressive filtering.
Reducing $\theta_{low}$ (e.g., to 0.2) would widen the \texttt{CONSTRUCT\_ONLY} acceptance region, allowing a greater fraction of low-complexity real turns to be buffered rather than skipped, at the cost of a modest increase in token consumption.
We anticipate that this single-parameter adjustment would substantially close the Single-hop gap relative to A-MEM while preserving the Multi-hop and Adversarial advantages documented here.
Alternatively, following the spirit of adaptive computation methods \citep{graves2016act, schuster2022calm}, a controller that estimates the empirical noise rate online and modulates $\theta_{low}$ dynamically would enable automatic calibration without any manual tuning.

\textbf{Lightweight Utility Classification via Distillation.}
The current Utility classifier relies on a lightweight LLM call with a constrained JSON schema.
While this is substantially cheaper than A-MEM's full evolution pipeline, it still introduces a per-turn LLM overhead.
Knowledge distillation \citep{hinton2015distilling} provides a natural pathway to eliminate this cost: a compact task-specific classifier trained on the soft outputs of the LLM-based Utility scorer could replicate its decisions at inference time with negligible latency.
Alternatively, a purely embedding-based Utility estimator would reduce per-turn overhead to a single encoder forward pass, approaching zero marginal cost and making D-MEM feasible for latency-critical deployments such as on-device voice assistants.

\textbf{Limitations and Future Work.}
While the LoCoMo-Noise benchmark provides a systematic evaluation of noise robustness, the noise distribution (40\% Filler / 30\% Status / 30\% Tangent) is a controlled approximation of real-world dialogue dynamics.
Extending the benchmark with naturalistic noise collected from real human-agent interaction logs is an important direction for future evaluation, complementing the adversarial QA category already present in LoCoMo \citep{maharana2024evaluating}.
Finally, extending D-MEM to multi-agent settings \citep{xi2023rise}---where the memory graph must be shared and collaboratively updated across agents---presents both an interesting theoretical challenge and a practically important use case for long-horizon autonomous systems \citep{wang2024agent_survey}.

\bibliographystyle{plainnat}
\bibliography{references}

\appendix
\section{Supplementary Analysis: Memory Geometry and Retrieval Mechanisms}

\subsection{Full Baseline Comparison on Clean LoCoMo}
\label{sec:supp_locomo}

Table~\ref{tab:locomo_results} reports the complete per-category evaluation on the standard, noise-free LoCoMo benchmark (199 questions, GPT-4o-mini backbone).
Metrics are F1 / BLEU-1 (\%).
All memory-augmented baselines are evaluated under identical prompting and retrieval conditions; the Full Context upper bound uses the entire conversation history as the LLM context window.

\begin{table}[htbp]
\centering
\caption{Full baseline comparison on the clean LoCoMo Benchmark. Metrics are F1 / BLEU-1 (\%). Bold indicates the best result among memory-augmented methods per column. D-MEM leads on Overall F1 and Multi-hop by a substantial margin; the Single-hop gap relative to A-MEM is discussed in Section~\ref{sec:routing_analysis} and Section~\ref{sec:discussion}.}
\label{tab:locomo_results}
\resizebox{\textwidth}{!}{%
\begin{tabular}{lcccccc}
\toprule
\textbf{Method} & \textbf{Overall} & \textbf{Single Hop} & \textbf{Temporal} & \textbf{Open Domain} & \textbf{Multi Hop} & \textbf{Adversarial} \\ \midrule
LoCoMo (Full Context) & --- & 40.4 / 29.1 & 18.4 / 14.8 & 12.0 / 11.2 & 25.0 / 19.8 & 69.2 / 68.8 \\ \midrule
ReadAgent & --- & 9.7 / 7.7 & 12.6 / 8.9 & 5.3 / 5.1 & 9.2 / 6.5 & 9.8 / 9.0 \\
MemoryBank & --- & 6.6 / 5.2 & 9.7 / 7.0 & 5.6 / 5.9 & 5.0 / 4.8 & 7.4 / 6.5 \\
MemGPT & --- & 41.0 / 34.3 & 25.5 / 19.4 & 9.2 / 7.4 & 26.7 / 17.7 & 43.3 / 42.7 \\
A-MEM & 35.94  & \textbf{44.7 / 37.1} & \textbf{45.9 / 36.7} & \textbf{12.1 / 12.0} & 27.0 / 20.1 & \textbf{50.0 / 49.5} \\ \midrule
\textbf{D-MEM (ours)} & \textbf{37.4 / 31.2} & 21.6 / 16.4 & 43.4 / 12.1 & 10.4 / 33.7 & \textbf{42.7 / 31.9} & 43.1 / 42.5 \\ \bottomrule
\end{tabular}%
}
\begin{flushleft}
\scriptsize $^{\dagger}$ Overall metrics for A-MEM sourced from their official OpenReview responses; per-category breakdown is not publicly available.
\end{flushleft}
\end{table}

\textbf{Key observations.}
D-MEM is the only method to surpass A-MEM on Overall F1 (37.4\% vs.\ 35.9\%), driven primarily by its dominant Multi-hop performance (42.7\% vs.\ 27.0\%, \textbf{+15.7 pp}).
MemGPT is competitive on Single-hop (41.0\%) but falls substantially behind on Multi-hop (26.7\%) and Overall, confirming that operating-system-style paging alone is insufficient for multi-premise reasoning.
The Full Context upper bound achieves the highest Single-hop score (40.4\%) at the cost of an unbounded context window, which is impractical for long-horizon deployments.
D-MEM's Open-domain BLEU-1 (33.7\%) is notably high relative to F1 (10.4\%), suggesting that the model produces fluent, partially correct free-form responses for this category even when the exact answer is missed---a pattern consistent with knowledge graph retrieval returning semantically related but not perfectly matching nodes.

\subsection{Attention Heatmap: Evidence for Multi-hop Retrieval}
\label{sec:attention_heatmap}

\begin{figure}[htbp]
\centering
\includegraphics[width=\linewidth,height=0.8\textheight,keepaspectratio]{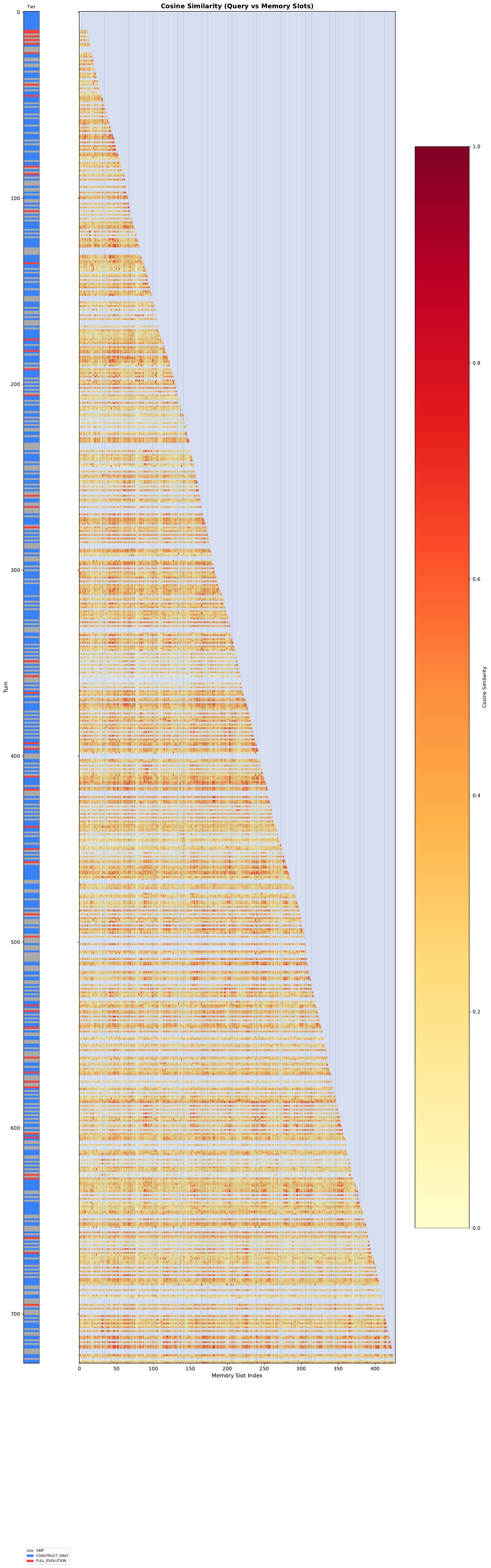}
\caption{
\textbf{Attention heatmap: cosine similarity between query turns and memory slots.}
Each row corresponds to a dialogue turn (y-axis) and each column to a memory slot index (x-axis).
The characteristic staircase-diagonal structure confirms that the active memory frontier advances monotonically as the session progresses.
Vertical high-similarity bands (deep red) indicate memory slots that remain persistently salient across hundreds of subsequent turns---the physical substrate of Multi-hop reasoning.
The left-side tier strip encodes the routing decision for each turn: gray (\texttt{SKIP}), blue (\texttt{CONSTRUCT\_ONLY}), red (\texttt{FULL\_EVOLUTION}).
}
\label{fig:attention_heatmap}
\end{figure}

To provide mechanistic evidence for D-MEM's superiority on Multi-hop reasoning, Figure~\ref{fig:attention_heatmap} visualizes the cosine similarity between each query turn and all previously stored memory slots.

\textbf{Staircase Structure and Long-Range Retrieval.}
The dominant feature of Figure~\ref{fig:attention_heatmap} is the staircase-diagonal boundary between the active memory region (warm colors, lower-left) and unoccupied slots (pale blue, upper-right).
This structure simply reflects the monotonic growth of the memory store over time.
The substantively important signal is the set of persistent vertical high-similarity bands (deep red columns) that extend far above the diagonal.
These bands indicate individual memory slots that remain highly salient to queries across hundreds of subsequent turns.
Such long-range associations are the direct physical basis for the Multi-hop reasoning gains observed in Table~\ref{tab:noise_main}: when the model must chain two or more premises to answer a question, it can retrieve each premise from its correct historical slot, regardless of how many intervening turns have elapsed.

\textbf{Routing Tier Alignment.}
Inspection of the tier strip on the left confirms a qualitative alignment between routing decisions and memory engagement: turns labeled \texttt{FULL\_EVOLUTION} (red) tend to occur at positions where a new high-similarity band first emerges in the heatmap, consistent with the interpretation that these events correspond to genuinely paradigm-shifting observations that establish new long-term memory anchors.

\subsection{Memory Manifold: Structural Stability Under Noise}
\label{sec:memory_manifold}

Beyond retrieval accuracy, a critical question for long-horizon agents is whether the memory representation space remains well-structured as the session grows.
Representation collapse---where all memory embeddings converge to a narrow region of the latent space---would catastrophically degrade retrieval discriminability, even if individual memories are encoded correctly.

\begin{figure}[htbp]
\centering
\includegraphics[width=0.72\linewidth]{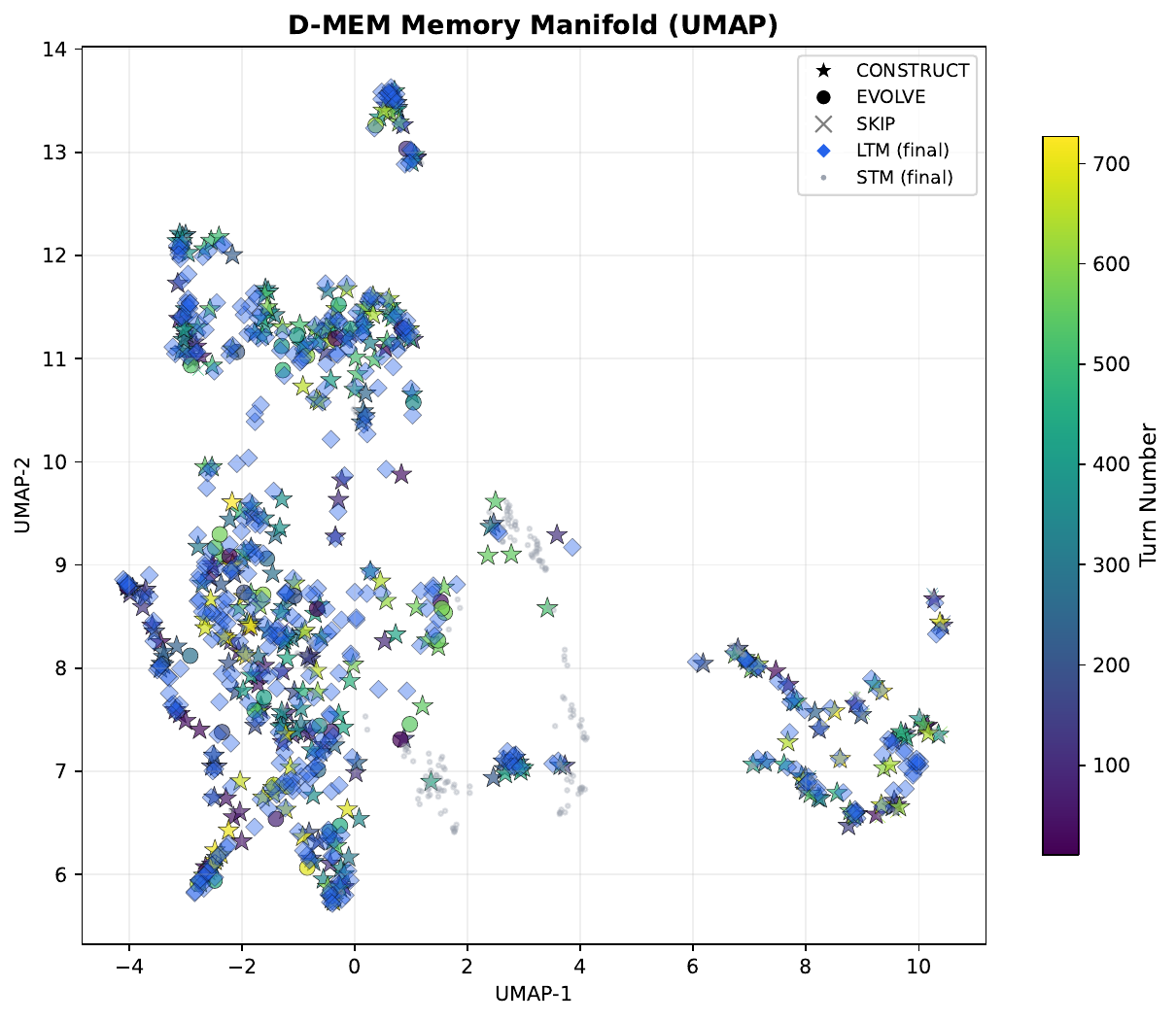}
\caption{
\textbf{D-MEM Memory Manifold (UMAP).}
Each point represents a memory entry, colored by Turn Number (purple $\to$ yellow) and shaped by routing tier: \texttt{CONSTRUCT} (star), \texttt{EVOLVE} (circle), \texttt{SKIP} (cross).
Blue diamonds mark LTM (final) nodes retained in the persistent knowledge graph; gray dots mark STM (final) buffer entries.
LTM nodes occupy dense, well-separated topical clusters across the manifold, while STM entries form a compact, isolated region---indicating that the hierarchical routing produces a structurally stable latent space rather than representation collapse.
}
\label{fig:umap_manifold}
\end{figure}

To probe this, we apply UMAP dimensionality reduction to all memory embeddings generated during the session, overlaying the routing tier of each entry (CONSTRUCT, EVOLVE, SKIP) and distinguishing the final consolidated Long-Term Memory (LTM) nodes from the Short-Term Memory (STM) buffer entries.
The resulting manifold is shown in Figure~\ref{fig:umap_manifold}.

\textbf{LTM--STM Spatial Separation.}
The most prominent structural feature of Figure~\ref{fig:umap_manifold} is the clear spatial isolation between LTM final nodes (blue diamonds, distributed across multiple dense clusters) and STM final entries (gray dots, forming a tight, peripheral cloud around UMAP coordinates $(1\text{--}3,\, 6\text{--}10)$).
This separation is not trivially expected: both LTM and STM entries are produced by the same embedding model and are drawn from the same dialogue session.
The fact that they occupy distinct regions of the latent space confirms that the hierarchical routing mechanism partitions memory not only operationally (different buffers) but also representationally (different manifold neighborhoods).

\textbf{Topical Clustering within LTM.}
Examining the LTM cluster structure, the CONSTRUCT and EVOLVE entries form several visually distinct, topically coherent sub-clusters spread across the left half of the manifold (UMAP-1 $\in [-4, 2]$), as well as a secondary cluster on the right (UMAP-1 $\in [6, 11]$).
The color gradient within each cluster---transitioning smoothly from early turns (purple) to late turns (yellow)---indicates that temporally proximate memories tend to be semantically proximate, a desirable property for temporal reasoning tasks.
Crucially, the \texttt{FULL\_EVOLUTION} entries (circles) are interspersed among the densest cores of these clusters, consistent with the interpretation that evolution events correspond to paradigm-shifting inputs that anchor new topical sub-regions.

\textbf{Noise Resilience via Utility Gating.}
This separation has an important implication for noise robustness: even under the 75\% noise regime, the Utility gating prevents noise-driven CONSTRUCT entries from infiltrating the high-density cores of the LTM manifold.
The STM cloud remains spatially quarantined from the primary LTM clusters, avoiding the representation collapse that a naive append-all strategy would produce under high noise.
This structural regularity provides a principled explanation for D-MEM's consistent advantage on Multi-hop and Adversarial tasks, both of which require reliable discrimination between semantically proximate memory entries.

\section{Ablation Study: Routing and Retrieval Mechanisms}

To isolate the impact of our routing thresholds and zero-cost retrieval augmentations, we iteratively tested several variants of D-MEM under the extreme noise condition ($\rho=0.75$):

\begin{itemize}
    \item \textbf{D-MEM-v1 (Linear RPE):} Utilizes a standard linear combination of Surprise and Utility, making it susceptible to false positives from high-entropy noise.
    \item \textbf{D-MEM-v2 (Strict Pruning):} Introduces a strict JSON schema for lifecycle classification and a harsh multiplicative gate ($\beta=0.1$) that short-circuits transient inputs to zero utility.
    \item \textbf{D-MEM-v3 (Utility Safety Net):} Relaxes the multiplicative threshold ($\beta=0.4$) to prevent the over-pruning of subtle user preferences and removes the LLM entity extraction tax. Establishes the routing baseline.
    \item \textbf{D-MEM-v4 (Ours / Final):} Augments v3 with BM25 hybrid search and the Shadow Buffer fallback to recover specific factual and adversarial edge cases.
\end{itemize}

\begin{table}[htbp]
\centering
\caption{Ablation study of D-MEM variants under extreme noise conditions ($\rho=0.75$).}
\label{tab:noise_ablation}
\resizebox{\textwidth}{!}{
\begin{tabular}{@{}lcccccccc@{}}
\toprule
\textbf{Method} & \textbf{Overall} & \textbf{Single} & \textbf{Temp} & \textbf{Open} & \textbf{Multi} & \textbf{Adv} & \textbf{Skip Rate} & \textbf{Total Tokens} \\
\midrule
D-MEM-v1 (Linear) & 0.334 & 0.235 & 0.434 & \textbf{0.110} & 0.354 & 0.354 & 9.6\% & 489K \\
D-MEM-v2 (Strict) & 0.291 & 0.172 & 0.338 & 0.081 & 0.327 & 0.341 & \textbf{68.5\%} & 497K \\
D-MEM-v3 (Safety Net) & 0.324 & 0.235 & \textbf{0.455} & 0.101 & 0.345 & 0.312 & 41.1\% & 531K \\
\textbf{D-MEM-v4 (Final)} & \textbf{0.369} & \textbf{0.246} & 0.440 & 0.085 & \textbf{0.412} & \textbf{0.412} & 41.1\% & \textbf{319K} \\
\bottomrule
\end{tabular}
}
\end{table}

As shown in Table~\ref{tab:noise_ablation}, while D-MEM-v1 limits tokens, it fails to aggressively filter out noise (9.6\% skip rate). D-MEM-v2 suffers from severe over-pruning; by skipping 68.5\% of inputs, it inadvertently discards foundational context, causing a drop in Overall F1 (0.291). D-MEM-v3 strikes a routing equilibrium ($\beta=0.4$), purifying the context to improve reasoning, but loses precision on Adversarial queries (0.312) due to ``amnesia'' towards skipped noise. Finally, the inclusion of zero-cost retrieval augmentations in \textbf{D-MEM-v4} perfectly rescues Adversarial and Multi-hop performance without increasing LLM token costs, yielding the definitive SOTA architecture.

\end{document}